\documentclass[journal=jctcce,manuscript=article]{achemso}

\usepackage[version=3]{mhchem} 
\usepackage{mathptmx}
\usepackage{multirow}
\usepackage{hyperref}
\usepackage{braket}
\usepackage{rotating}
\usepackage{color}
\usepackage{float}
\usepackage{adjustbox}
\usepackage{amsfonts}

\DeclareMathOperator{\Tr}{Tr}

\newcommand{\corr}[1]{#1}

\newcommand{\fig}{Fig.}

\newcommand{\figref}[1]{\fig~\ref{#1}}

\newcommand{\cm}{\ensuremath{\text{cm}^{-1}}}

\newcommand{\Eqref}[1]{Equation~(\ref{#1})}

\newcommand{\secref}[1]{Section~\ref{#1}}

\author{Viktor Zaverkin}
\author{Julia Netz}
\author{Fabian Zills}
\author{Andreas Köhn}
\author{Johannes Kästner}
\email{kaestner@theochem.uni-stuttgart.de}

\affiliation{Institute for Theoretical Chemistry, University of Stuttgart, Pfaffenwaldring 55, 70569 Stuttgart, Germany}
\title[Thermally Averaged Magnetic Anisotropy Tensors via Machine Learning Based on Gaussian Moments]{Thermally Averaged Magnetic Anisotropy Tensors via Machine Learning Based on Gaussian Moments}

\abbreviations{IR,NMR,UV}

\begin{document}

\begin{tocentry}

\includegraphics[width=\linewidth]{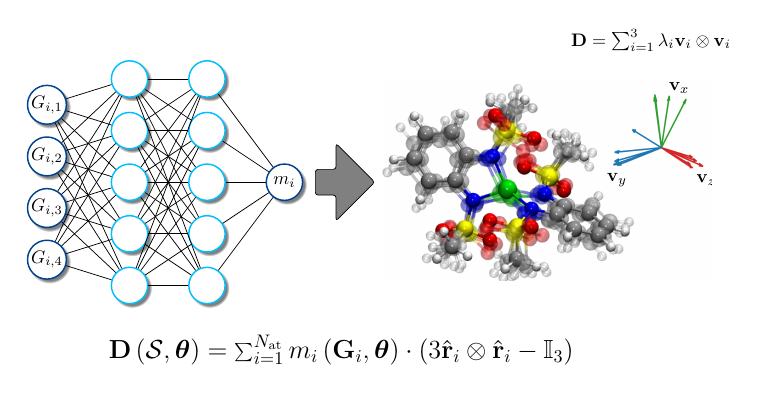}

\end{tocentry}

\begin{abstract}
We propose a machine learning method to model molecular tensorial quantities, namely the magnetic anisotropy tensor, based on the Gaussian-moment neural-network approach. We demonstrate that the proposed methodology can achieve an accuracy of 0.3--0.4~$\cm$ and has excellent generalization capability for out-of-sample configurations. Moreover, in combination with machine-learned interatomic potential energies based on Gaussian moments, \corr{our approach can be applied to study the dynamic behavior of magnetic anisotropy tensors and provide a unique insight into spin-phonon relaxation.}
\end{abstract}

\section{\label{sec:intro} Introduction}

There is an ongoing interest in the investigation of magnetic properties of transition metal complexes and their possible applications as single-molecule magnets (SMMs), molecular quantum bits, and spintronic devices.\cite{Gatteschi2006, Craig2015, GaitaArino2019, Sanvito2011} One of the most important factors influencing the magnetic properties is the magnetic anisotropy in the ground spin state.\cite{Sessoli2003, Craig2015} 
Tailoring promising complexes in a way to achieve a large barrier for magnetic relaxation and minimizing other decay pathways\cite{Rechkemmer2016, Harman2010, Rogez2005, Bamberger_2021} requires a detailed understanding of the underlying principles that determine the structure-property relationships.

While sampling based on first-principles methods can provide some insight, it is usually limited by its high computational cost. Approximate methods allow for simulations even taking into account the large size of conformational and chemical space of interest. Machine learning approaches with their ability to learn any complex non-linear relationship between a structure and its related property and with their generalization capability are perfect candidates for the specific task of modeling the magnetic anisotropy.
During the last decades, machine learning methods have been gaining importance in several fields in computational chemistry allowing for, e.g., the construction of machine-learned force fields and prediction of vibrational spectra.\cite{Blank1995, Lorenz2004, Lorenz2006, Behler2007, Behler2011, Behler2016, Shapeev2016, Gastegger2017, Gubaev2018, Dral2018, Zhang2018, Yao2018, Westermayr2019, Unke2019, Zaverkin2020, Molpeceres2020, Dral2020, Molpeceres2021, Christensen2020, Faber2018, Zaverkin2021}
The ability of neural networks (NNs) to interpolate any non-linear functional relationship~\cite{Hornik1991} promoted their broad application in computational chemistry and materials science. NNs were initially applied to represent potential energy surfaces (PESs) of small atomistic systems\cite{Blank1995, Lorenz2004} and were later extended to high-dimensional systems.\cite{Behler2007} Once trained, the computational cost of machine-learned potentials (MLPs) based on NNs is independent of the number of data points used for training.

Case studies of the application of machine learning algorithms to study the magnetic properties of molecules and materials are somewhat rare. Recently, an approach for the construction of machine-learned interatomic potentials which are capable of reproducing both vibrational and magnetic degrees of freedom was proposed\cite{Novikov2021} and applied to bcc iron. Closer to the present work are the investigations presented in Refs.~\citenum{Lunghi2020, Lunghi2020_2}, where tensorial properties such as the zero-field splitting (ZFS) tensor $\mathbf{D}$ and the  Zeeman-splitting (ZS) tensor $\mathbf{g}$ were modeled by machine learning approaches.

In this work, we build upon the methodology developed in our group for constructing efficient and accurate interatomic potentials, referred to as Gaussian moment neural network (GM-NN).~\cite{Zaverkin2020, Zaverkin2021}
The GM-NN uses neural networks (NNs) to map novel symmetry-preserving local atomic descriptors, Gaussian moments (GMs), to auxiliary atomic quantities and includes both the geometric and alchemical information about the atomic species of both the central and neighbor atoms. For all atomic contributions, only a single NN has to be trained, in contrast to using an individual NN for each species as frequently done in the literature.\cite{Behler2007, Behler2016} To allow for efficient training on properties such as $\mathbf{D}$ tensors we introduce a novel approach for the encoding of relevant invariances into the output of the respective machine learning model. Moreover, we propose a neural network architecture for the task of modeling the magnetic anisotropy of SMMs.

To assess the quality of the surrogate models obtained employing the proposed approach we thoroughly benchmark the predictive accuracy on three promising SMMs: \ce{[Co(N2S2O4C8H10)2]}$^{2-}$,~\cite{Rechkemmer2016} \ce{[Fe(tpa)$^{Ph}$]}$^{-}$,~\cite{Harman2010, Atanasov2011} and \ce{[Ni(HIM2-py)2NO3]}$^{+}$ complexes.~\cite{Rogez2005} Finally, we use the \ce{[Co(N2S2O4C8H10)2]}$^{2-}$ complex as a case study to investigate the applicability of the proposed approach to the dynamics of the $\mathbf{D}$ tensor and \corr{examine} its ability to model spin-phonon coupling effects.

\corr{In this work, we employ the open-source package Gaussian moment neural network (GM-NN) for constructing MLPs and applying them in atomistic simulations. The GM-NN source code is available free-of-charge from \href{https://gitlab.com/zaverkin\_v/gmnn}{gitlab.com/zaverkin\_v/gmnn}.}

\section{\label{sec:methods} Machine-Learning Models}

This section first introduces the \corr{representation of an atomistic system used to encode molecular and solid-state structures suitable for machine learning (ML) models. Second, we describe the} approach to machine learning modeling the magnetic anisotropy tensors including the neural network architecture and its training, based on our previous work on Gaussian moments.~\cite{Zaverkin2020, Zaverkin2021}
\corr{Finally}, we \corr{outline} the construction of the machine-learned interatomic potential energy used in \secref{sec:md_results} for molecular dynamics simulations of \ce{[Co(N2S2O4C8H10)2]}$^{2-}$.

\subsection{\label{sec:representation} Molecular Representation}

\corr{A molecular or solid-state system is defined by its Cartesian coordinates $\mathbf{r}_i \in \mathbb{R}^3$ and its atomic numbers $Z_i$ which, in the following discussion, are combined to $\mathcal{S} = \{\mathbf{r}_i, Z_i\}_{i=1}^{N_\mathrm{at}}$ for simplicity.
In the community of neural network (NN) model chemistry, it is usual to approximate properties by a sum of their atomic contributions. For example, the total energy of a system can be approximated\cite{Behler2007} by a sum of atomic energies $\mathcal{E}_i$
\begin{equation}
\mathcal{E} \left( \mathcal{S}, \boldsymbol{\theta} \right) = \sum_{i=1}^{N_\mathrm{at}} \mathcal{E}_i\left(\mathbf{G}_i, \boldsymbol{\theta}\right).
\end{equation}
Total charge and the dipole moments can be defined via the atomic point charges $q_i$
\begin{equation}
\begin{split}
& Q_\mathrm{tot}\left( \mathcal{S}, \boldsymbol{\theta} \right) = \sum_{i=1}^{N_\mathrm{at}} q_i\left(\mathbf{G}_i, \boldsymbol{\theta}\right),\\
& \boldsymbol{\mu}\left( \mathcal{S}, \boldsymbol{\theta} \right) = \sum_{i=1}^{N_\mathrm{at}} q_i\left(\mathbf{G}_i, \boldsymbol{\theta}\right) \mathbf{r}_{i},
\end{split}
\end{equation}
where each $\mathcal{E}_i$ and $q_i$ are given as the output of an artificial NN, and all relevant invariances are encoded in the local representation $\mathbf{G}_i(\mathcal{S}, \boldsymbol{\beta})$ with $\boldsymbol{\beta}$ being trainable parameters. Thus, the main purpose of machine learning is to find such parameters $\boldsymbol{\theta}$ that the mapping $f: \mathcal{S} \mapsto \{E, Q_\mathrm{tot}, \boldsymbol{\mu}, \dots\}$ is as close as possible to the reference values, $\{E^\mathrm{ref}, Q_\mathrm{tot}^\mathrm{ref}, \boldsymbol{\mu}^\mathrm{ref}, \dots\}$.}

\corr{The most challenging aspect of a machine learning model applied to a physicochemical problem is the definition of an appropriate representation of an atomistic system. In this work, we employ Gaussian moment representation which defines the local atomic environment as\cite{Zaverkin2020, Zaverkin2021}
\begin{equation}
\label{eq:psi}
    \begin{split}
        \boldsymbol{\Psi}_{i, L, s} = \sum_{j \neq i} R_{Z_i, Z_j, s}\left(r_{ij}, \boldsymbol{\beta}\right)  \hat{\mathbf{r}}_{ij}^{\otimes L},
    \end{split}
\end{equation}
where we use the radial and angular components of the atomic distance vector $\mathbf{r}_{ij} = \mathbf{r}_i - \mathbf{r}_j$, i.e. $r_{ij} = \lVert\mathbf{r}_{ij}\rVert_2$ and $\hat{\mathbf{r}}_{ij} = \mathbf{r}_{ij} / r_{ij}$, as inputs to nonlinear radial functions $R_{Z_i, Z_j, s}\left(r_{ij}, \boldsymbol{\beta}\right)$ with trainable parameters $\boldsymbol{\beta}$ and the $L$-fold tensor product $\hat{\mathbf{r}}_{ij}^{\otimes L} = \hat{\mathbf{r}}_{ij} \otimes \cdots \otimes \hat{\mathbf{r}}_{ij}$, respectively. $Z_i$ and $Z_j$ correspond to the nuclear charges of the central atom $i$ and its atomic neighbors $j$. As nonlinear radial functions we employ a weighted sum of Gaussian functions, rescaled by the cosine cutoff function,~\cite{Behler2007} see Ref.~\citenum{Zaverkin2020, Zaverkin2021}. The parameters $\boldsymbol{\beta}$ are optimized during training similar to other parameters of an atomistic NN by minimization of a loss function. A rotationally invariant representation is obtained by computing full tensor contractions as described in Ref.~\citenum{Zaverkin2020, Zaverkin2021}.
}

\corr{As a final remark on the representation of an atomistic system, we want to discuss the possibility of learning different physicochemical properties by a single ML model. In general, it should be possible, given that the corresponding properties are available for the same atomistic system. Unfortunately, in this work, the energy and atomic forces are calculated for the full periodic system while the magnetic anisotropy tensors are computed for isolated molecules cut from them, see \figref{fig:1}. That implies that both atomistic systems provide different sets of input features and, therefore, cannot directly be combined. For this reason, in the following sections we describe the construction of two separate ML models for learning $\mathbf{D}$ tensors and potential energy surfaces.}

\subsection{\label{sec:zfs_ml} The $\mathbf{D}$ Tensor}

Mediated by spin-orbit coupling, the components of spin multiplets (with spin quantum numbers $S\ge 1$) split characteristically even in the absence of an external magnetic field (zero-field splitting, ZFS). This effect is usually described by a phenomenological spin Hamiltonian\cite{AbragamBleaney}
\begin{equation}
    H_\mathrm{ZFS}=\hat{\mathbf{S}}\cdot \mathbf{D}\cdot \hat{\mathbf{S}}
\end{equation}
where $\hat{\mathbf{S}}$ is a (pseudo) spin operator and \textbf{D} is a $3 \times 3$ symmetric, traceless tensor, usually called ZFS tensor or \textbf{D} tensor. The anisotropy parameters $D$ and $E$ (often simply referred to as $D$ and $E$ values) can be then obtained from the eigenvalues of $\mathbf{D}$ ($X$, $Y$, and $Z$) as
\begin{equation}
    \begin{split}
    & D = \frac{3}{2} Z,\\
    & E = \frac{1}{2}(X-Y),
    \end{split}
\end{equation}
where $Z$ corresponds to the eigenvalue of $\mathbf{D}$ with largest absolute value. The normalized eigenvectors of \textbf{D} are the anisotropy axes of the system.

\subsubsection{Incorporation of Symmetries into the ML Framework}

The aim of this work is to predict $\mathbf{D}$ tensors using artificial neural networks (NNs). The $\mathbf{D}$ tensor is a \textbf{(1)} traceless \textbf{(2)} symmetric tensor, which implies that $\Tr \mathbf{D} = \sum_i D_{ii} = 0$ and $D_{ij} = D_{ji}$. Additionally, it \textbf{(3)} transforms under rotation as $\tilde{\mathbf{D}} = \mathbf{R}\mathbf{D}\mathbf{R}^T$, where $\mathbf{R}$ is an orthogonal matrix, but is \textbf{(4)} invariant to translations. These properties have to be encoded into the respective machine learning model to allow for efficient training. Moreover, unlike the energy, the $\mathbf{D}$ tensor cannot be reduced to a natural scalar atomic property which makes its prediction more difficult using the atomistic neural networks.

To the best of our knowledge, there are only two studies in the literature of modeling tensorial properties such as $\mathbf{g}$ and $\mathbf{D}$ tensors.\cite{Lunghi2020, Lunghi2020_2} In Ref.~\citenum{Lunghi2020}, where ridge regression was used to fit $\mathbf{D}$ tensors, the rotational equivariance was imposed by requiring the regression coefficients to transform as spherical tensors, i.e. employing the sum over the Wigner matrices corresponding to rigid rotations. In Ref.~\citenum{Lunghi2020_2}, where artificial neural networks were used to fit $\mathbf{g}$ tensors, a rotation operator $\Lambda$ was defined by, e.g., the Kabsh algorithm to impose the desired symmetries. In this work, we present a more rigorous and efficient way of encoding the symmetries \textbf{(1)}--\textbf{(4)} into machine learning algorithms.

We approach the modeling of zero-field splitting tensors by introducing a fictitious atomic quantity $m_i$ assigned to each atom $i$. A tensor which satisfies \textbf{(1)--(3)} can be defined using the tensor product of Cartesian vectors, similar to the quadrupole moment, employing the aforementioned atomic quantity $m_i$
\begin{equation}
\label{eq:zfs_1}
\mathbf{D}\left( \mathcal{S}, \boldsymbol{\theta} \right) = \sum_{i=1}^{N_\mathrm{at}} m_i\left(\mathbf{G}_i, \boldsymbol{\theta}\right)\cdot\left(3\mathbf{r}_i\otimes\mathbf{r}_i - \|\mathbf{r}_{i}\|_2^2\mathbb{I}_{3}\right)~,
\end{equation}
where $\otimes$ is the tensor product, $\|\mathbf{r}_{i}\|_2$ is the length of the respective Cartesian vector, and $\mathbb{I}_3$ is a $3\times 3$ identity matrix.

Unfortunately, the above expression violates translational invariance \textbf{(4)} of the respective tensorial property. This issue can be resolved by shifting the coordinate system by, e.g., $\bar{\mathbf{r}} = 1/N_\mathrm{at} \sum_{i} \mathbf{r}_i$, which is system dependent. Note that the shift can be defined by any other procedure since its aim is only to impose the translational invariance \textbf{(4)}. However, we should emphasize that we have found that using the pairwise distances $\mathbf{r}_{ij} = \mathbf{r}_i - \mathbf{r}_j$, where $i$ can be selected to be the metal atom, could be disadvantageous in terms of predictive accuracy due to missing contributions from the central atom. 

Using the respective shift vector $\bar{\mathbf{r}}$ results in a new position vector $\mathbf{r}_i \rightarrow \mathbf{r}_i - \bar{\mathbf{r}}$. Additionally, we have found that it is advantageous to use the normalized vectors $\hat{\mathbf{r}}_i = \mathbf{r}_i / \|\mathbf{r}_{i}\|_2$. In total, the resulting expression reads
\begin{equation}
\label{eq:zfs_2}
\mathbf{D}\left( \mathcal{S}, \boldsymbol{\theta} \right) = \sum_{i=1}^{N_\mathrm{at}} m_i\left(\mathbf{G}_i, \boldsymbol{\theta}\right)\cdot\left(3\hat{\mathbf{r}}_i\otimes\hat{\mathbf{r}}_i - \mathbb{I}_{3}\right)~,
\end{equation}
where $m_i$ is predicted by an atomistic NN.

As a final remark we want to point out that any symmetry of a tensorial property $\mathbf{P}$ can be modelled by the procedure proposed above if one defines $\mathbf{P} = \sum_i m_i \mathbf{A}_i$, where $\mathbf{A}_i$ is a tensor satisfying the symmetry of $\mathbf{P}$ and $m_i$ is a machine-learned scalar value. In our example, we define $\mathbf{A}_i = 3\hat{\mathbf{r}}_i\otimes\hat{\mathbf{r}}_i - \mathbb{I}_{3}$ to impose \textbf{(1)}--\textbf{(4)}, but if the respective property is, e.g, not traceless one could use $\mathbf{A}_i = \hat{\mathbf{r}}_i\otimes\hat{\mathbf{r}}_i$.

\subsubsection{Network Architecture and Training} \label{sec:d_net}

We use a fully-connected feed-forward neural network consisting of two hidden layers of the following functional form\corr{, similar to our previous work,~\cite{Zaverkin2021}}
\begin{equation}
\begin{split}
    y_i\left(\mathbf{G}_i, \boldsymbol{\theta}\right) &= 0.1 \cdot \mathbf{b}^{(3)} + \frac{1}{\sqrt{d_2}} \mathbf{W}^{(3)} \phi\left(0.1 \cdot \mathbf{b}^{(2)} + \right. \\ & \quad \left. \frac{1}{\sqrt{d_1}} \mathbf{W}^{(2)} \phi \left(0.1 \cdot \mathbf{b}^{(1)} + \frac{1}{\sqrt{d_0}} \mathbf{W}^{(1)} \mathbf{G}_i\right)\right)~,
\end{split}
\end{equation}
where $\mathbf{W}^{(l)}$ and $\mathbf{b}^{(l)}$ are weights and biases of the respective layer $l$. 
As an input to the neural network, $\mathbf{G}_i$, we use the recently proposed trainable local invariant representation based on Gaussian moments.\cite{Zaverkin2020, Zaverkin2021}
The parameters $0.1$ and $1/\sqrt{d_l}$ correspond to the so-called NTK parameterisation.\cite{Jacot2018} We initialize weights of the fully-connected part by drawing the respective entries from a normal distribution with zero mean and unit variance. The trainable bias vectors are initialized to zero. As an activation function we use the Swish/SiLU activation function\cite{Hendrycks2016, Elfwing2018, Ramachandran2018} $\phi(x) = \alpha x/\left(1 + \exp\left(-x\right)\right)$ multiplied by a scalar $\alpha$. We choose $\alpha \approx 1.6765$ such that $\mathbb{E}_{x \sim \mathcal{N}\left(0, 1\right)} \phi(x)^2 = 1$, i.e., the activation function preserves the second moment if the input is standard Gaussian.~\cite{Klambauer2017, Arora2019, Lu2020}

In order to aid the training process, the output of the neural network can be scaled and shifted by the standard deviation $\sigma$ and the mean $\mu$ of the reference $\mathbf{D}^\mathrm{ref}$ tensor values, similar to the scaling and shifting the atomic energy output.~\cite{Zaverkin2020, Zaverkin2021}
Note that we used for the computation of $\sigma$ and $\mu$ only those elements of $D_{ij}$ which satisfy $i \geq j$ to avoid double counting and excluded one diagonal element since the respective tensor is traceless. The convergence of the model can be improved even further by making these parameters trainable as well as dependent on the atomic species, i.e. $\sigma_{Z_i}$ and $\mu_{Z_i}$. The final output of the network reads
\begin{equation}
    m_i\left(\mathbf{G}_i, \boldsymbol{\theta}\right) = y_i\left(\mathbf{G}_i, \boldsymbol{\theta}\right) \sigma_{Z_i} + \mu_{Z_i}~.
\end{equation}

To train the neural network on reference values for $\mathbf{D}^\mathrm{ref}$ tensors, we minimize the following loss function
\begin{equation}
\label{eq:loss}
\begin{split}
\mathcal{L}_{\mathbf{D}}\left(\boldsymbol{\theta}\right) = \sum_{k=1}^{N_\mathrm{Train}} & \lVert \mathbf{D}_k^\mathrm{ref} - \mathbf{D}(\mathcal{S}_k, \boldsymbol{\theta})\rVert_2^2~,
\end{split}
\end{equation}
to optimize the respective parameters of the trainable representation, fully-connected neural network part as well as the parameters which scale and shift the output of the neural network. Note that we train only on those elements of $\left(\mathbf{D}\right)_{ij}$ tensor which satisfy $i \geq j$ and we define $\mathbf{D}^\mathrm{ref}$ and $\mathbf{D}$ as
\begin{equation}
    \begin{split}
        & \mathbf{D}^\mathrm{ref} = \Big(D_{11}^\mathrm{ref}, D_{12}^\mathrm{ref}, D_{13}^\mathrm{ref}, D_{22}^\mathrm{ref}, D_{23}^\mathrm{ref}, D_{33}^\mathrm{ref}\Big)^T \\
        & \mathbf{D} = \Big(D_{11}, D_{12}, D_{13}, D_{22}, D_{23}, -\left(D_{11} + D_{22}\right)\Big)^T~.
    \end{split}
\end{equation}

To minimize the loss function in \Eqref{eq:loss} the Adam optimizer~\cite{Adam2015} with hyper-parameters $\beta_1 = 0.9$, $\beta_2 = 0.999$, $\epsilon = 10^{-7}$, and a mini-batch of 32 structures is employed.
Moreover, we allow for layer-wise learning rates which decay linearly to zero by multiplying them with $(1-r)$ where $r = \mathrm{step}/\mathrm{max\_step}$. Overall, we use an initial learning rate of $0.02$ for the parameters of the fully connected layers, $0.02$ for the parameters of the trainable GM representation, $0.025$ and $0.025$ for the shift and scale parameters.

Throughout this work we used an architecture with an input dimension of $d_0 = 360$, two hidden layers with $d_1 = d_2 = 512$ hidden neurons, and an output layer which has a single $d_3 = 1$ output neuron. Each model in \secref{sec:results} is trained for 1000 epochs. Overfitting was prevented by using the early stopping technique.~\cite{Prechelt2012} After each epoch, the mean absolute errors (MAE) of the $\mathbf{D}$ tensor was evaluated on the validation set. After training, the model with the minimal MAE on the validation set was selected for further applications. While the selected hyper-parameters worked reasonably well on the selected systems for us, we want to emphasize that other trade-offs between the number of training epochs and the initial learning rates can be achieved.

Note that to compute machine-learned $\mathbf{D}$ tensors during an MD simulation we interfaced our approach with the ASE package (v. 3.21.0).~\cite{Hjorth2017} For tracking the accuracy we employed the query-by-committee (QbC) approach~\cite{Settles2009} during MD simulations. For this purpose, we trained a committee of 3 models on the same split of the data set but using randomly initialized parameters and reported the obtained uncertainty
\begin{equation}
\label{eq:uncertainty}
    \sigma_\mathrm{ens}\left(\mathcal{S}\right) = \sqrt{\frac{1}{N_\mathrm{ens}}\sum_{i=1}^{N_\mathrm{ens}} \left(y_i\left(\mathcal{S}\right) - \bar{y}\left(\mathcal{S}\right)\right)^2}~,
\end{equation}
where $N_\mathrm{ens}$ is the number of models in the committee, i.e. $N_\mathrm{ens} =3$. $\bar{y}\left(\mathcal{S}\right) = 1/N_\mathrm{ens}\sum_{i=1}^{N_\mathrm{ens}} y_i\left(\mathcal{S}\right)$ is the mean of the property prediction (energy, atomic force element, or $\mathbf{D}$-tensor element, respectively) over the committee. All models were trained within the Tensorflow framework~\cite{Abadi2015} on an NVIDIA Tesla V100-SXM-32GB GPU. \corr{The training of an ensemble of 3 models for 1000 epochs took from 4 min (100 \ce{[Co(N2S2O4C8H10)2]}$^{2-}$ structures) to 3 h (2900 \ce{[Ni(HIM2-py)2NO3]}$^{+}$ structures).}

\subsection{\label{sec:mlip} Potential Energy}

Thermal averaging of magnetic anisotropy tensors requires an interatomic potential that fulfills two main premises. The underlying model has to produce sufficiently accurate atomic forces for molecular dynamics (MD) simulations, i.e., comparable to the level of theory employed for the generation of the training data, and allow for efficient computations, i.e., comparable to empirical force fields. Machine-learning algorithms have found a broad application in computational chemistry since they satisfy both conditions provided a suitable molecular descriptor that encodes structural and alchemical information. 

As for the $\mathbf{D}$ tensors, we employed the Gaussian moment neural network (GM-NN) approach\cite{Zaverkin2020, Zaverkin2021}
to construct  the PES. Again, we use a neural network with an input dimension of $d_0 = 360$, two hidden layers with $d_1 = d_2 = 512$ hidden neurons, and an output layer which has a single $d_3 = 1$ output neuron.

To train the GM-NN model the combined loss function 
\begin{equation}
\label{eq:loss_pes}
\begin{split}
\mathcal{L}_{\mathcal{E}, \mathbf{F}}\left(\boldsymbol{\theta}\right) = \sum_{k=1}^{N_\mathrm{Train}} & \Bigg[\lambda_\mathcal{E} \lVert \mathcal{E}_k^\mathrm{ref} - \mathcal{E}(\mathcal{S}_k, \boldsymbol{\theta})\rVert_2^2 +  \\ & \quad  \frac{\lambda_F}{3N_\mathrm{at}^{(k)}} \sum_{i=1}^{N_\mathrm{at}^{(k)}} \lVert \mathbf{F}_{i,k}^\mathrm{ref} - \mathbf{F}_i\left(\mathcal{S}_k, \boldsymbol{\theta}\right)\rVert_2^2\Bigg]~,
\end{split}
\end{equation}
is minimized. Here, $N_\mathrm{at}^{(k)}$ is the number of atoms in the respective structure. The reference values for the energy and atomic force are denoted by $\mathcal{E}_k^\mathrm{ref}$ and $\mathbf{F}_{i,k}^\mathrm{ref}$, respectively. The parameters $\lambda_\mathcal{E}$ and $\lambda_F$ were set to $1~\text{au}$ and $12N_\mathrm{at}^{(k)}$~au~{\AA}$^2$, respectively. The network was trained using the Adam optimizer \citep{Adam2015} with $32$ molecules per mini-batch. The layer-wise learning rate was set to $0.03$ for the parameters of the fully connected layers, $0.02$ for the trainable representation, 0.05 and 0.001 for the shift and scale parameters of atomic energies respectively. All learning rates we allowed to decay linearly to zero. \corr{For more information, see Ref.~\citenum{Zaverkin2021}.}

To run MD simulations with the machine-learned potentials (MLPs) we interfaced the GM-NN approach with the ASE package (v. 3.21.0).~\cite{Hjorth2017} For tracking the accuracy of MLPs we employed the query-by-committee (QbC) approach~\cite{Settles2009} during MD simulations. For this purpose, we trained a committee of 3 models on the same split of the data set but using randomly initialized parameters and reported on the attained uncertainty, see \Eqref{eq:uncertainty}. All models were trained within the Tensorflow framework~\cite{Abadi2015} on an NVIDIA Tesla V100-SXM-32GB GPU. \corr{The training of an ensemble of 3 models for 1000 epochs took about 40 h (3100 \ce{[Co(N2S2O4C8H10)2]}$^{2-}$ periodic structures).}

\section{\label{sec:data} Test Systems and Computational Details}

In this section, we describe the generation of the data used to construct machine-learned interatomic potentials as well as machine learning models for $\mathbf{D}$ anisotropy tensors and an overview of selected test systems.

\subsection{\label{sec:systems} Test System Description}

\begin{figure*}
    \begin{minipage}{0.3\textwidth}
    \centering
    \includegraphics[width=5cm]{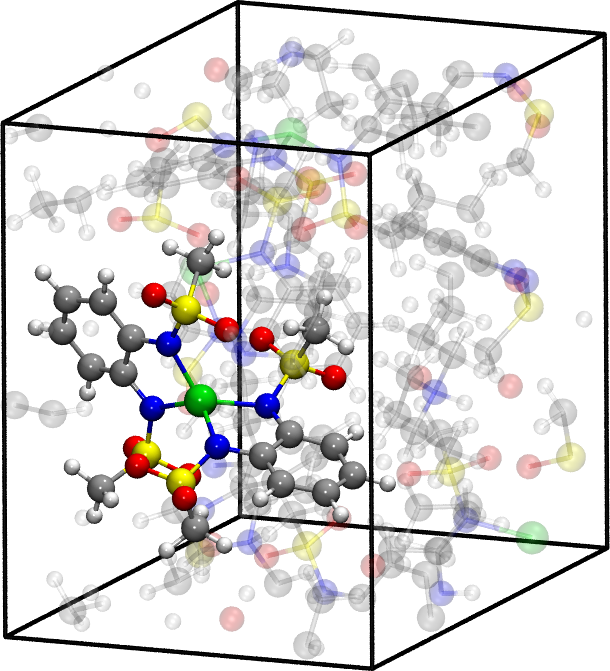}
    \end{minipage}\hfill
    \begin{minipage}{0.3\textwidth}
    \centering
    \includegraphics[width=4cm]{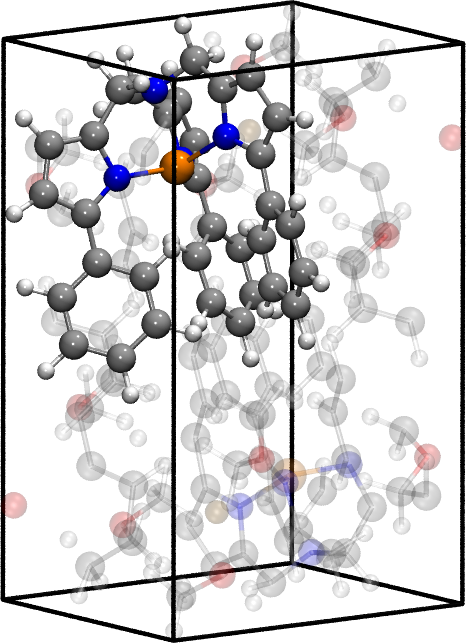}
    \end{minipage}\hfill
    \begin{minipage}{0.3\textwidth}
    \centering
    \includegraphics[width=5cm]{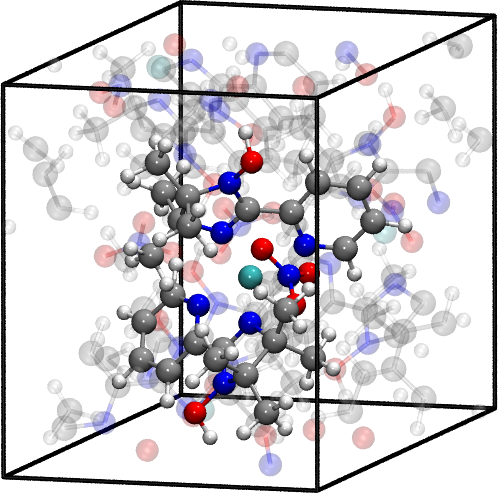}
    \end{minipage}
    \caption{Illustration of the structure of (left) the \ce{[Co(N2S2O4C8H10)2]}$^{2-}$ unit cell, (middle) the \ce{[Fe(tpa)$^{Ph}$]}$^{-}$ unit cell, and (right) the \ce{[Ni(HIM2-py)2NO3]}$^{+}$ unit cell. The color code is: Co green, Fe orange, Ni turquoise, Na brown, S yellow, C grey, O red, N blue, and H white.}
    \label{fig:1}
\end{figure*}

We selected the following three promising candidates for SMMs (in fact even single-ion magnets, SIMs): \ce{[Co(N2S2O4C8H10)2]}$^{2-}$,~\cite{Rechkemmer2016} \ce{[Fe(tpa)$^{Ph}$]}$^{-}$,~\cite{Harman2010, Atanasov2011} and \ce{[Ni(HIM2-py)2NO3]}$^{+}$.~\cite{Rogez2005} \figref{fig:1} illustrates the unit cells of the systems and the corresponding complexes used in the cluster models.

\paragraph*{\textbf{\ce{[Co(N2S2O4C8H10)2]}$^{2-}$}.}
One of the most promising high-anisotropy SIMs has been reported by Rechkemmer \textit{et al.}~\cite{Rechkemmer2016} It has a single cobalt ion center which is bound to two doubly deprotonated 1,2-bis(methanesulfonamido)benzene ligands resulting in a distorted tetrahedral coordination sphere. In the crystal, the charge is compensated by two \ce{NHEt3+} cations. The zero-field splitting parameter $D^\mathrm{exp}=-115\pm 20~\cm$ has been experimentally determined by fitting to AC and DC susceptibility and magnetic hysteresis measurements.~\cite{Rechkemmer2016}

\paragraph*{\textbf{\ce{[Fe(tpa)$^{Ph}$]}$^{-}$}.}
This is one example out of a big family of similar complexes which all feature an iron center ion in trigonal-pyramidal surrounding of a pyrrolide ligand tpa$^\text{R}$.~\cite{Harman2010} The complex studied here has phenyl attached to the main pyrrolide ligand (R=Ph). The counter ion in the crystal is \ce{Na+(H3COC2H4OCH3)3}. Experimentally the magnetic parameters have been determined to be $D^\mathrm{exp}=-26\pm2~\cm$, $E^\mathrm{exp}=5~\cm$ by AC and DC magnetic susceptibility measurements and fitting procedures.~\cite{Harman2010} This family of complexes was also theoretically studied by Atanasov \textit{et al.} on the level of complete active space self-consistent field (CASSCF) wave functions in conjunction with N-electron valence perturbation theory (NEVPT2) and quasi-degenerate perturbation theory (QDPT).~\cite{Atanasov2011}

\paragraph*{\textbf{\ce{[Ni(HIM2-py)2NO3]}$^{+}$}.}
One example of a nickel complex with a high magnetic anisotropy has been synthesized by Rogez \textit{et al.}~\cite{Rogez2005}. The nickel ion is influenced by a highly distorted octahedral coordination sphere consisting of two bidentate ligands HIM2-py as well as an O,O’-chelating nitrate ligand. In the crystal, the positive charge is compensated by a second \ce{NO3-} ion without direct contact to the nickel ion.  Measurements of the magnetization versus field, HF-HFEPR, and FDMRS resulted in fitted magnetic parameters $D^\mathrm{exp}=-10.1 \pm 0.1~\cm$ and $E^\mathrm{exp}=0.202 \pm 0.01~\cm$.\cite{Rogez2005}

\subsection{Data Generated via AIMD} \label{sec:aimd_gen}

For all training data sets an ab initio molecular dynamics (AIMD) calculation has been carried out using the PAW method \cite{Blochl1994_2, Kresse1999} with the PBE functional, \cite{Perdew1997} as implemented in the VASP program package.\cite{Kresse1993, Kresse1996, Kresse1996_2} A Hubbard $U$ correction term has been used with the simplified (rotationally invariant) approach introduced by Dudarev \emph{et al.}\cite{Dudarev1998} The used values of $U$ are 3.3 eV for Co, 4.0 eV for Fe and 6.4 eV for Ni.\citep{Mann2016} Dispersion corrections were applied by the zero damping DFT-D3 method.\cite{Grimme2010} 

As a starting point, the crystal structure was first optimized using an energy cutoff for the plane-wave basis of 600 eV and the 'accurate precision' settings in VASP, the projection operators were evaluated in real-space. The Brillouin zone was sampled by a Monkhorst-Pack grid (Co system: 2$\times$2$\times$2; Fe system: 3$\times$3$\times$2; Ni system: 4$\times$4$\times$2). The stress tensor was calculated and all degrees of freedom were allowed to change in relaxation.

After that, we performed an AIMD simulation with VASP, using the same settings as before but with a cutoff energy of 400~eV and the 'normal precision' settings. Only the $\Gamma$ point was considered. For each system shown in \figref{fig:1} we calculated 5000 time steps of 1~fs for the temperatures 100K, 300K, 400K, 450K, and 500K each using a Nos\'e{}--Hoover thermostat as implemented in VASP.

We chose more structures from the MD runs at the higher temperatures, i.e. 200, 600, 800, 900, and 1000 sample structures randomly chosen from the MD runs at 100K, 300K, 400K, 450K, and 500K, respectively. This was done because more diverse conformations are visited at higher temperature and therefore dynamics performed at a higher temperature contains more information necessary for the construction of reliable interatomic potentials.

\subsection{D-Tensor Data} \label{sec:d_gen}

To generate reference values for the $\mathbf{D}$ tensor on which models were subsequently trained, we cut the corresponding molecules from the periodic structures of the AIMD simulation, as illustrated in \figref{fig:1}. Effects of the neighboring molecules and the crystal field on the $\mathbf{D}$ tensor have thus been neglected for the present study. In total, we obtained 3500 configurations for each test system, for which magnetic properties were calculated using the Molpro program package.\cite{MOLPRO_brief} The orbitals were optimized by the configuration-averaged Hartree-Fock (CAHF) procedure\cite{McWeeny:Book, McWeeny:MP28-1273, Hallmen:JCP147-164101, Calvello2018} using the Karlsruhe def2-SVP basis sets.\cite{Weigend:PCCP7-3297} The active space consisted of the five d orbitals in each of the systems (see below). Based on these orbitals, a complete-active space configuration interaction (CASCI) calculation was performed to obtain the spin-free states. These were afterwards used in a spin-orbit configuration interaction (SO-CI) calculation, using a  mean-field spin-orbit operator\cite{Marian1996, Hess1996, Berning2000}, based on the CAHF average density for the mean field.

The following setup is used for each of the test systems: For \ce{[Co(N2S2O4C8H10)2]}$^{2-}$ we chose a CAS(7,5) and the SO-CI calculation was carried out using 40 doublet and 10 quartet CASCI states; for \ce{[Fe(tpa)$^{Ph}$]}$^{-}$ we chose a CAS(6,5) and the SO-CI calculation was based on 50 singlet, 45 triplet, and 5 quintet CASCI states; for \ce{[Ni(HIM2-py)2NO3]}$^{+}$ we finally chose a CAS(8,5) and the SO-CI calculation used 15 singlet and 10 triplet CASCI states.

From the SO-CI calculations, the  $\mathbf{D}$ tensor was extracted using the pseudo-spin procedure described by Chibotaru and Ungur.~\cite{Chibotaru2012}

\section{\label{sec:results}Results and Discussion}

Here we apply the proposed approach for machine learning symmetric traceless tensors, namely $\mathbf{D}$ tensors, to systems described in \secref{sec:data}. For the example of \ce{[Co(N2S2O4C8H10)2]}$^{2-}$ we demonstrate the applicability of our approach to studying relevant magnetic properties of SMMs.

\subsection{\label{sec:ml_results} Machine Learning of the D Tensor}

\begin{figure*}
    \centering
    \includegraphics[width=16cm]{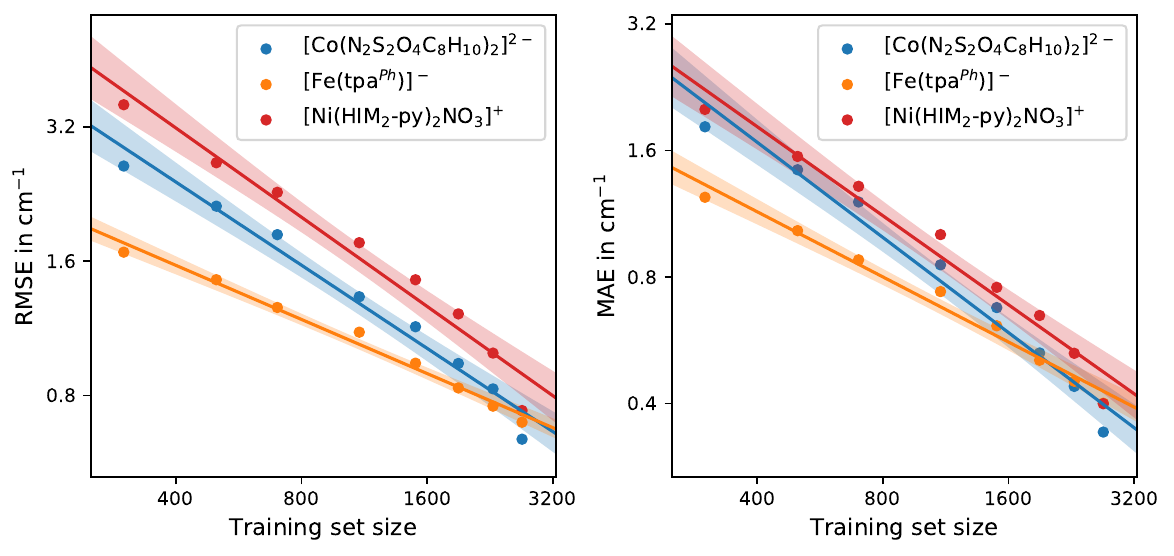}
    \caption{Learning curves for the \ce{[Co(N2S2O4C8H10)2]}$^{2-}$, \ce{[Fe(tpa)$^{Ph}$]}$^{-}$, and \ce{[Ni(HIM2-py)2NO3]}$^{+}$ data sets. The root-mean-squared errors (RMSE) and the mean absolute errors (MAE) of the $\mathbf{D}$-tensor components are plotted against the training set size. Linear fits are displayed for clarity and shaded areas denote the 95~\% confidence intervals for linear regression.}
    \label{fig:2}
\end{figure*}

To study the performance of the proposed approach for machine learning magnetic anisotropy tensors, or specifically $\mathbf{D}$ tensors, we trained our model on \ce{[Co(N2S2O4C8H10)2]}$^{2-}$, \ce{[Fe(tpa)$^{Ph}$]}$^{-}$, and \ce{[Ni(HIM2-py)2NO3]}$^{+}$ data. Each data set contained structures and $\mathbf{D}$ tensor elements of 3500 configurations. In \figref{fig:2} we report the mean absolute error (MAE) and the root-mean-square error (RMSE) in $\cm$. The cutoff radius employed in the definition of invariant atomic representation was set to $8.0$~\AA. 
All other hyper parameters are described in \secref{sec:d_net}.

\figref{fig:2} reports the MAE and RMSE of $\mathbf{D}$ tensor predictions as a function of the number of training samples. For all training set sizes, we have randomly drawn 300 additional structures  as validation data to track overfitting during training. Since the validation data indirectly influence the selected set of trainable parameters, all values presented in \figref{fig:2} are obtained for the test data that have not been seen during training. For example, for 2900 training data we have used the remaining 300 structures to test the model, while for 1500 training data 1700 structures were used for the same purpose.

\begin{figure*}
        \centering
        \includegraphics[width=12cm]{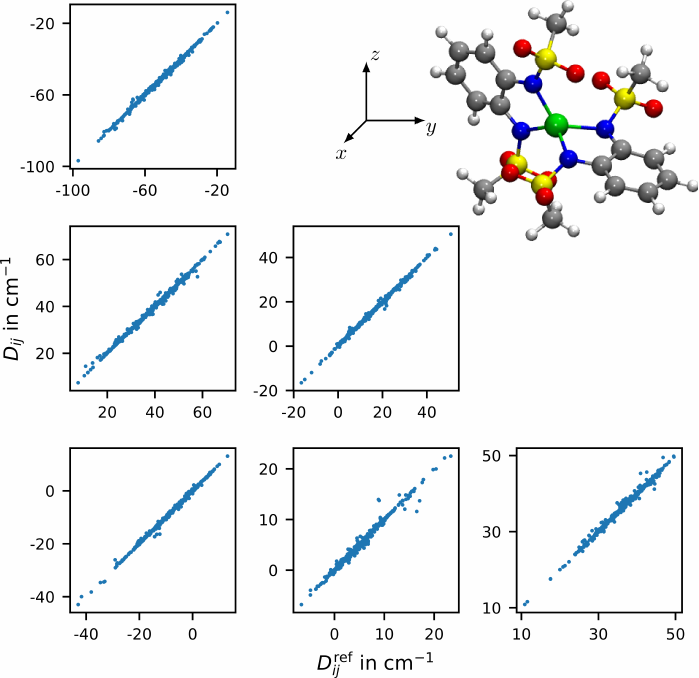}
    \caption{Correlation of the machine-learned symmetric elements ($ i \geq j $) of the zero-field splitting tensor ($D_{ij}$) with the corresponding reference values ($D_{ij}^\mathrm{ref}$) for all structures in the test \ce{[Co(N2S2O4C8H10)2]}$^{2-}$ data. For all predictions, the model trained on 2900 reference structures was used. The respective coordinate system of the periodic box as well as an example substructure for which $\mathbf{D}$ tensor was computed are shown as an inset.}
    \label{fig:3}
\end{figure*}

In general, we notice that the proposed approach leads to models that learn quite efficiently on the reference data. For example, the RMSE for the \ce{[Co(N2S2O4C8H10)2]}$^{2-}$ data set is reduced by a factor of two when doubling the training data set size. It can be observed that the RMSE for \ce{[Ni(HIM2-py)2NO3]}$^{+}$ is slightly higher compared to other systems, while \ce{[Fe(tpa)$^{Ph}$]}$^{-}$ shows lower RMSE values for smaller training data set sizes. The former observation can be explained by the higher flexibility of \ce{[Ni(HIM2-py)2NO3]}$^{+}$. This leads to a larger conformational space sampled during ab initio molecular dynamics (AIMD) in \secref{sec:aimd_gen} and as a result to a broader range of $\mathbf{D}$-tensor elements, see Fig.~S2. For \ce{[Fe(tpa)$^{Ph}$]}$^{-}$ the situation is different since structural differences lead only to a slight variation in $\mathbf{D}$-tensor elements, see Fig.~S1, compared to other systems, \figref{fig:3} and Fig.~S2.

A direct comparison to previous models dealing with machine learning magnetic anisotropy tensors, e.g., Refs.~\citenum{Lunghi2020, Lunghi2020_2}, is impossible since no standard benchmark is available in the literature, in contrast to, e.g., QM9~\cite{Ruddigkeit2012, Ramakrishnan2014} and MD17~\cite{Schuett2017_2, Chmiela2017, Chmiela2018} data sets for testing machine-learned interatomic potentials. Therefore, we use the presented data sets, available free-of-charge from Ref.~\citenum{Zaverkin2021b}, to benchmark the models developed for predicting $\mathbf{D}$ tensors.

In previous work,\cite{Lunghi2020} for \ce{[Co(N2S2O4C8H10)2]}$^{2-}$ an RMSE value of 2.2~$\cm$ was obtained when training on 900 structures, while we obtain 1.5~$\cm$. In Ref.~\citenum{Lunghi2020} the training data set was generated starting from an optimized structure in a vacuum and then displacing atoms by a maximum of $\pm 0.05$~\AA{} (500~structures), $\pm 0.1$~\AA{} (500~structures), and $\pm 0.2$~\AA{} (500~structures), while we extracted the conformations from an AIMD simulation. Thus, we expect that our data sets cover a broader conformational space, making the training more difficult (which would lead to higher MAE/RMSE values) but facilitating the prediction of properties for out-of-sample configurations.

In \figref{fig:3}, we show the correlation of the reference and machine-learned results of each $D_{ij}$ element in the $\mathbf{D}$ tensor for the \ce{[Co(N2S2O4C8H10)2]}$^{2-}$ system. We again use only those structures the machine learning model has not seen during training. From \figref{fig:3} we clearly see a perfect correlation between both values in accordance with the low MAE (RMSE) value of 0.30~$\cm$ (0.58~$\cm$), obtained by training the model on 2900 reference structures. Similar results were obtained for \ce{[Fe(tpa)$^{Ph}$]}$^{-}$ and \ce{[Ni(HIM2-py)2NO3]}$^{+}$, see Fig. S1 and Fig. S2. In total, taking into account the broadly sampled conformational space and an excellent out-of-sample predictive accuracy, our approach should be applicable to large-scale molecular dynamics simulations.

\subsection{\label{sec:md_results} Dynamics of the \textbf{D} Tensor}

We assess the quality of machine-learned $\mathbf{D}$ tensors by applying the machine learning methodology to the \ce{[Co(N2S2O4C8H10)2]}$^{2-}$ system to study its magnetic/physicochemical properties. While a detailed analysis is out of the scope of this paper, we provide merely a qualitative overview to demonstrate the broad applicability of our approach.

To analyze the properties of \ce{[Co(N2S2O4C8H10)2]}$^{2-}$ we ran molecular dynamics (MD) simulations in the canonical (NVT) statistical ensemble, carried out within the ASE simulation package\cite{Hjorth2017} using a Langevin thermostat at the temperatures of 25, 50, 75, and 100~K. All MD runs were performed over 2.5~ns using a time step of 0.5~fs. The atomic velocities were initialized with a Maxwell--Boltzmann distribution for the temperatures of 25, 50, 75, and 100~K, respectively.

Forces for molecular dynamics were generated by an ensemble of three machine-learned interatomic potentials, see \secref{sec:mlip}. To train the machine-learned interatomic potentials a cutoff radius of 6.5~\AA{} was employed. The ensembling technique provides us with an error estimate of the potential during simulation. We have found the machine-learned potential to be very accurate with the uncertainty between models ranging from $0.10 \pm 0.06$~kcal/mol/\AA{} to $0.15 \pm 0.10$~kcal/mol/\AA{}, see Fig. S3. For $\mathbf{D}$ tensor predictions we also use an ensemble of 3 models with an uncertainty ranging from $0.11 \pm 0.06~\cm$ to $0.20 \pm 0.11~\cm$, see Fig S3. These values allow us to claim that our machine-learned models are suitable for the following analysis.

\subsubsection{Dependence of the \textbf{D} Tensor on the Structure}

To study the structural dependence of the $\mathbf{D}$ tensor we evaluated the average \ce{Co-N} distance $d_{\ce{CoN}}$ and the tetrahedral order parameter $q_{\widehat{\ce{NCoN}}}$ as~\cite{Errington2001}
\begin{equation}
    \begin{split}
        & d_{\ce{CoN}} = \frac{1}{4} \sum_{i=1}^4 d_{i}~, \\
        & q_{\widehat{\ce{NCoN}}} = 1 - \frac{3}{8}\sum_{i=1}^3\sum_{j=i+1}^4 \left(\cos\left(\theta_{ij}\right) + \frac{1}{3}\right)^2~,
    \end{split}
\end{equation}
where $d_i$ and $\theta_{ij}$ are the distance and angle formed by the metal center and its neighboring nitrogen atoms. Note that in the literature usually the average angle between nitrogen atoms belonging to the same ligand is used as order parameter,~\cite{Lunghi2020} or its deviation from a perfect tetrahedral angle $T_d = 109.5^\circ$ is discussed.~\cite{Titi2013} We have employed the tetrahedral order parameter $q_{\widehat{\ce{NCoN}}}$ often used when studying the structure of liquid water.~\cite{Liu2018} This parameter contains the information about angular distribution but is rescaled in such a way that its value varies between 0 (if the arrangement of all atoms is random) and 1 (in a perfect tetrahedral network). Moreover, it has a value of 0.5 for a perfect quadratic planar configuration, which allows us to look into both possible configurations of the first coordination sphere of \ce{[Co(N2S2O4C8H10)2]}$^{2-}$, tetrahedral and quadratic planar. 

\begin{figure*}
    \centering
    \includegraphics[width=16cm]{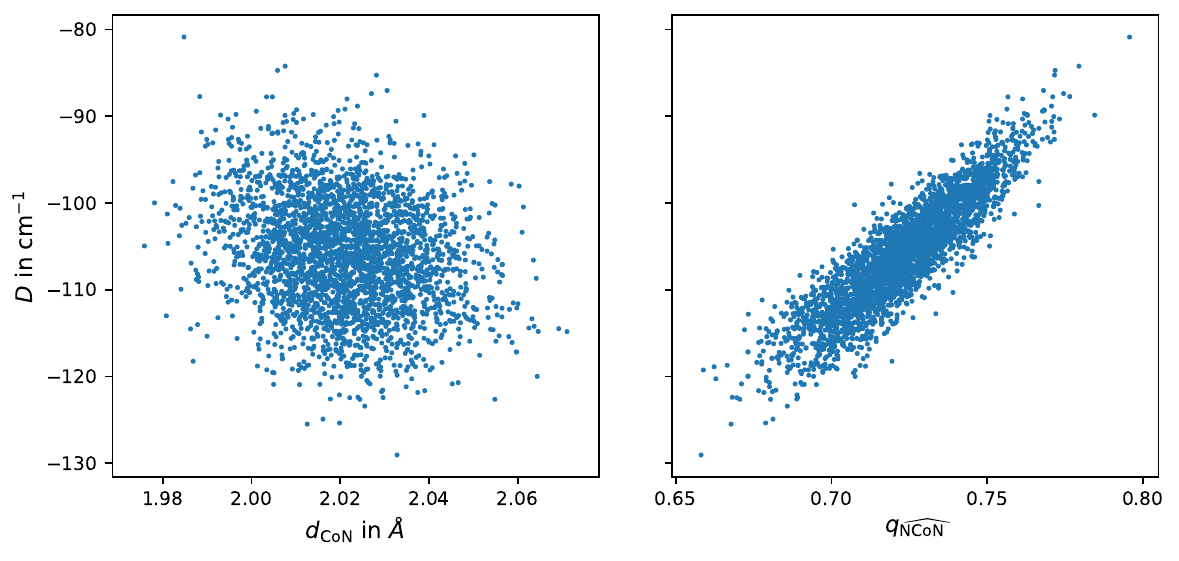}
    \caption{Correlation of the magnetic axial anisotropy $D = \tfrac{3}{2} Z$, where $Z$ is the largest eigenvalue of the $\mathbf{D}$ tensor, with (left) the average \ce{Co-N} distance $d_{\ce{CoN}}$ and (right) tetrahedral order parameter $q_{\widehat{\ce{NCoN}}}$. The latter yields a value of 1 for the perfect tetrahedral structure, 0.5 for a perfect quadratic planar structure, and 0 for a random mutual arrangement of central and neighboring atoms. The tetrahedral order parameter $q_{\widehat{\ce{NCoN}}}$ yields a linear correlation coefficient of 0.88, while the average \ce{Co-N} distance does not correlate with $D$ and yields a linear correlation coefficient of $-0.22$.}
    \label{fig:4}
\end{figure*}

\figref{fig:4} shows the correlation of the magnetic axial anisotropy $D = \tfrac{3}{2} Z$, where $Z$ is the eigenvalue of the $\mathbf{D}$ tensor with the largest absolute value, with the order parameters presented above. Data obtained from MD at 100~K are shown since they provide the broadest range of conformations and the broadest range of $D$ values. From \figref{fig:4} we see that the average \ce{Co-N} distance $d_{\ce{CoN}}$ correlates only marginally with the magnetic axial anisotropy $D$ for which we obtained a linear correlation coefficient of $-0.22$. This is in agreement with results found in Ref.~\citenum{Lunghi2020} taking into consideration that a minimal average \ce{Co-N} distance of about 1.98~\AA{} could be sampled at 100~K. Note that in Ref.~\citenum{Lunghi2020} a stronger correlation was found for distances less than 2.0~\AA{}. However, from our MD simulations we see that such configurations are too rare at 100~K to be taken into account.

\figref{fig:4} clearly shows a strong dependence of the magnetic axial anisotropy $D$ on the tetrahedral order parameter $q_{\widehat{\ce{NCoN}}}$ for which we obtained a linear correlation coefficient of 0.88. When changing the structure from tetrahedral coordination to a quadratic planar one, a strong increase of the magnetic anisotropy is observed. Our results are in perfect agreement with recent results that suggest that the $\widehat{\ce{NCoN}}$ angle represents the main path for improving \ce{Co}$^{2+}$ single-ion magnets.~\cite{Lunghi2020, Rechkemmer2016, Fataftah2014, Carl2015, Suturina2017, Wu2019}

\subsubsection{Thermal Distribution of the D Tensor}

\begin{figure*}
\begin{minipage}{0.6\textwidth}
\centering
\includegraphics[width=10cm]{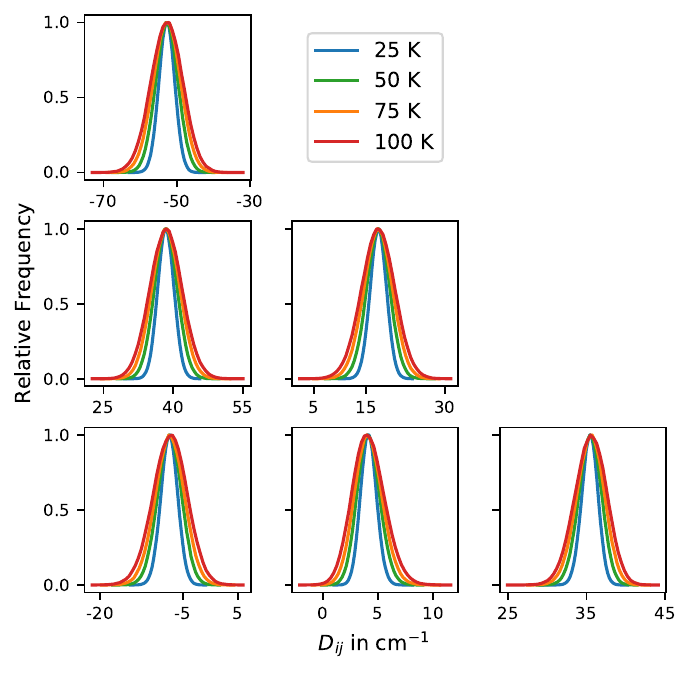}
\end{minipage}\hfill
\begin{minipage}{0.4\textwidth}
\centering
\includegraphics[width=6cm]{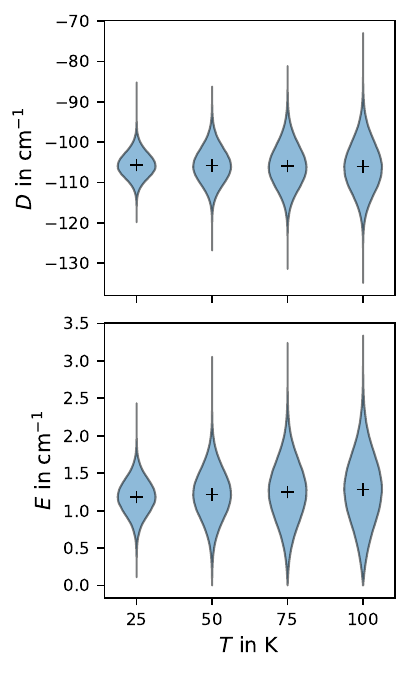}
\end{minipage}  
\caption{(Left) Thermal broadening of the elements $D_{ij}$ with $ i \geq j $ of the zero-field splitting tensor $\mathbf{D}$ obtained from sampling configurations over 2.5~ns long molecular dynamics. All values were computed employing the machine learning model described in \secref{sec:zfs_ml}. For all predictions, the model trained on 2900 reference structures was used. (Right) Temperature dependence of $D = \tfrac{3}{2} Z$ and $E =\left|X - Y\right|/2$, where $X$, $Y$, and $Z$ are the eigenvalues of the $\mathbf{D}$ tensor, with $Z$ being its largest absolute eigenvalue.}
\label{fig:5}
\end{figure*}

Besides the correlation of the magnetic axial anisotropy $D$ with structural order parameters $d_{\ce{CoN}}$ and  $d_{\widehat{\ce{NCoN}}}$ we study the variation in magnitude and orientation due to thermal fluctuations in the $\mathbf{D}$ anisotropy tensor. \figref{fig:5} (left) shows the distribution of the individual $D_{ij}$ components of the $\mathbf{D}$ tensor sampled over 2.5~ns. It can be seen that all elements $D_{ij}$ are approximately normally distributed with broader distributions for higher temperatures. The elements are symmetric with $D_{ij} = D_{ji}$, hence only $D_{ij}$ for $i \geq j$ are shown. It should be noted that we have found that the mean of the distribution of each element $D_{ij}$ remains almost unchanged and the maximal deviation equals 0.1~$\cm$. However, the distribution of each element $D_{ij}$ is broadened by approximately a factor of 2 when increasing the temperature from 25~K to 100~K.

In order to relate the dynamics of the magnetic axial anisotropy $D = \tfrac{3}{2} Z$ to the structural dynamics of the \ce{[Co(N2S2O4C8H10)2]}$^{2-}$ complex we display the respective distribution in \figref{fig:5} (right) along with the distribution for $E = \left|X - Y\right|/2$. Note that $X$, $Y$, and $Z$ are the eigenvalues of $\mathbf{D}$ with $Z$ being the one with the largest absolute value. We have found the values for $D$ to range from $-105.7 \pm 3.1~\cm$ to $-106.0 \pm 6.2~\cm$, i.e. the mean is almost temperature-independent while the standard deviation is doubled when increasing the temperature by a factor of 4 in accordance with results for $D_{ij}$. Note that our values estimated from MD simulations with machine learning are very close to the experimental one of $-115 \pm 20~\cm$.~\cite{Rechkemmer2016} The values for $E$ range from $1.2 \pm 0.2~\cm$ to $1.3 \pm 0.5~\cm$.

\subsubsection{Time Correlation Functions and Spin-Phonon Coupling}

Finally, and mainly as an outlook to future applications of our approach, we study the coupling of the $\mathbf{D}$ tensor of the \ce{[Co(N2S2O4C8H10)2]}$^{2-}$ complex to the dynamics of the periodic atomic structure (velocity vectors $\mathbf{v} = \dot{\mathbf{x}}$). For this purpose, we define time-dependent $\mathbf{D}$ tensor and velocity autocorrelation functions (ACFs) as
\begin{equation}
    \begin{split}
        & C_{\mathbf{D}\mathbf{D}} \left( t \right) = \frac{1}{6} \sum_{i=1}^{3}\sum_{j \geq i}^3 \frac{\langle D_{ij}\left(0\right) D_{ij}\left(t\right)\rangle}{\langle D_{ij}\left(0\right) D_{ij}\left(0\right)\rangle}~,\\
        & C_{\mathbf{v}\mathbf{v}} \left( t \right) = \frac{1}{3 N_\mathrm{at}} \sum_{i=1}^{N_\mathrm{at}}\sum_{j =1}^3 \frac{\langle v_{ij}\left(0\right) v_{ij}\left(t\right)\rangle}{\langle v_{ij}\left(0\right) v_{ij}\left(0\right)\rangle}~,
    \end{split}
\end{equation}
respectively. Using these ACFs it is possible to compute the corresponding spectra by performing a Fourier transform or by employing the maximum-entropy approach.~\cite{Burg1975, Koen2013, Martini2021} In this work we compute the power spectral density function $S_\omega$ of interest employing the \texttt{memspectrum} package.~\cite{Martini2021_2}

\begin{figure*}
\begin{minipage}{0.4\textwidth}
\centering
\includegraphics[width=6cm]{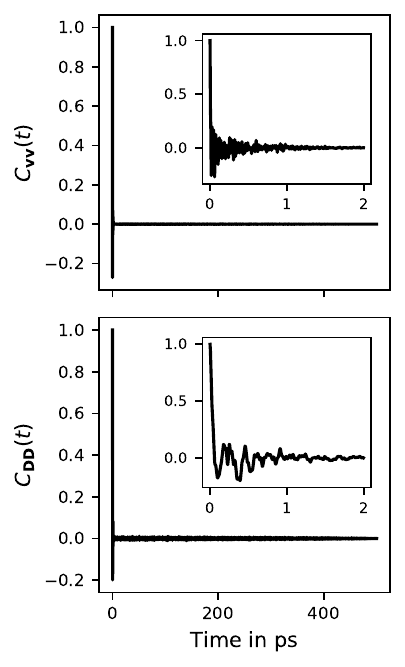}
\end{minipage}\hfill
\begin{minipage}{0.6\textwidth}
\centering
\includegraphics[width=10cm]{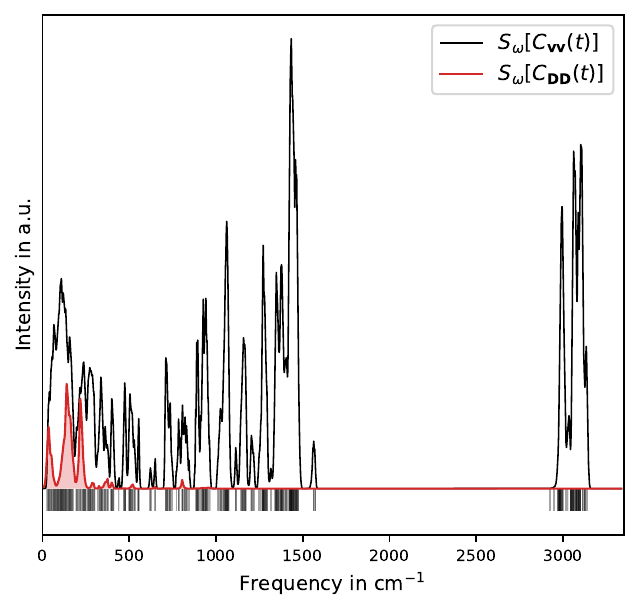}
\end{minipage}
\caption{(Left) Velocity and $\mathbf{D}$ tensor autocorrelation functions, $C_{\mathbf{v}\mathbf{v}} \left( t \right)$ and $C_{\mathbf{D}\mathbf{D}} \left( t \right)$ respectively, obtained by averaging over 5 independent MD trajectories at 50~K of a length of 500~ps. (Right) Power spectral density function $S_{\omega}$ of the velocity and $\mathbf{D}$ tensor autocorrelation functions. The gray vertical lines at the bottom of the graph correspond to harmonic frequencies.}
\label{fig:6}
\end{figure*}

\figref{fig:6} shows the velocity and $\mathbf{D}$ tensor ACFs (left) as well as the respective power spectral density functions $S_{\omega}[C_{\mathbf{v}\mathbf{v}} \left( t \right)]$ and $S_{\omega}[C_{\mathbf{D}\mathbf{D}} \left( t \right)]$ (right). From the latter, in principle, the spin-phonon coupling coefficients $\partial D / \partial q_\alpha$ beyond the harmonic approximation can be calculated, which provide the interaction strength between the spin and the atomic movements.\cite{Lunghi2017} The present study is restricted to the demonstration of the capability of the proposed approach, a detailed analysis will be presented in future work. For now, we computed the spectra for the $\Gamma$-point only, while the inclusion of the full Brillouin zone ($k$-dependence) may be important to estimate the spin lifetime. Even with our preliminary spectra, we are able to deduce similar conclusions as previous work\cite{Lunghi2017} found for \ce{[Fe(tpa)$^{Ph}$]}$^{-}$ in that the most important vibrations in the spin-phonon relaxation process are the low-energy ones. They are predominantly populated under typical experimental conditions.

\section{\label{sec:conclusion} Conclusion}

In this work, we have presented a machine learning approach based on Gaussian moments~\cite{Zaverkin2020, Zaverkin2021}
for tensorial properties on the example of the zero-field splitting ($\mathbf{D}$) tensor. It enables an efficient prediction and modeling of magnetic properties of single-molecule magnets. The presented approach was extensively tested on three systems \ce{[Co(N2S2O4C8H10)2]}$^{2-}$, \ce{[Fe(tpa)$^{Ph}$]}$^{-}$, and \ce{[Ni(HIM2-py)2NO3]}$^{+}$ in terms of its predictive accuracy and its reliability during real-time simulations.

Training the proposed model on the respective reference $\mathbf{D}$ values for the \ce{[Co(N2S2O4C8H10)2]}$^{2-}$ we have observed an improved accuracy compared to previous studies,~\cite{Lunghi2020} especially taking into account the larger sampled configurational space. In total, for all systems tested in this work we could achieve an MAE of 0.3--0.4~$\cm$ and an RMSE of 0.6--0.7~$\cm$ for the models trained on 2900 reference structures. Moreover, we have shown that the proposed approach, once trained on a sufficiently large configurational space, is able to predict $\mathbf{D}$ tensor values for millions of conformations not seen before with a negligibly small uncertainty of 0.11--0.20~$\cm$ obtained by employing the query-by-committee approach. This demonstrates the excellent generalization capability of our approach.

In combination with machine-learned interatomic potentials we were able to run 2.5~ns long molecular dynamics simulations at temperatures of 25, 50, 75, and 100~K. Using the respective trajectories we could analyze several properties of the \ce{[Co(N2S2O4C8H10)2]}$^{2-}$ complex taken as an example. Analysing the dependence of the magnetic axial anisotropy on average \ce{Co-N} distance and an $\widehat{\ce{NCoN}}$ angle-dependent order parameter $q_{\widehat{\ce{NCoN}}}$ we have found that the $\widehat{\ce{NCoN}}$ angle represents the main path for improving \ce{Co}$^{2+}$ single-ion magnets, in accordance with recent results.~\cite{Lunghi2020, Rechkemmer2016, Fataftah2014, Carl2015, Suturina2017, Wu2019} Moreover, we could estimate the thermal average of $D$ and $E$ values. For the former, a very good agreement with the experiment has been found, while for the latter no experimental data are available.

Besides the structure, we investigated the dynamical behavior of the $\mathbf{D}$ tensor of the \ce{[Co(N2S2O4C8H10)2]}$^{2-}$ complex. Even in such  preliminary work, we observed the expected behavior that the low-energy vibrations are important for the spin-phonon relaxation process.

In summary, our developments aim to provide an alternative way for the efficient modeling of magnetic properties of single molecular magnets via machine learning. While the current setup is based on a relatively simple complete-active space configuration interaction treatment of the molecular properties, our approach can be easily extended to more elaborate methods, e.g. by using transfer learning. Future work will furthermore deal with an application of the proposed methodology to allow for a detailed analysis of spin-phonon relaxation processes.

\begin{acknowledgement}

We thank the Deutsche Forschungsgemeinschaft (DFG, German Research Foundation) for supporting this work by funding EXC 2075 - 390740016 under Germany's Excellence Strategy and through grant no INST 40/575-1 FUGG (JUSTUS 2 cluster). We acknowledge the support by the Stuttgart Center for Simulation Science (SimTech). We also like to acknowledge the support by the state of Baden-Württemberg through the bwHPC consortium for providing computer time. V. Z. acknowledges financial support received in the form of a PhD scholarship from the Studienstiftung  des  Deutschen  Volkes (German National Academic Foundation).

\end{acknowledgement}

\bibliography{main}

\end{document}